# An Oracle Strongly Separating Deterministic Time from Nondeterministic Time, via Kolmogorov Complexity

David Doty*

**Abstract**

Hartmanis used Kolmogorov complexity to provide an alternate proof of the classical result of Baker, Gill, and Solovay that there is an oracle relative to which P is not NP. We refine the technique to strengthen the result, constructing an oracle relative to which a conjecture of Lipton is false.

## 1 Introduction

Hartmanis [3] used time-bounded Kolmogorov complexity to provide an alternate proof of the classical result of Baker, Gill, and Solovay [1] that there is an oracle separating P and NP.

We strengthen the result to obtain a stronger separation between deterministic time classes and nondeterministic time classes. Lipton [5, 6] has conjectured that it may be possible to simulate an arbitrary nondeterministic Turing machine running in time $t(n)$ in much fewer than $2^{t(n)}$ steps; for instance, $1.1^{t(n)}$. Our result implies that this conjecture, if true, requires non-relativizing techniques to prove.

## 2 Construction of the Oracle

We use the notation and definitions of complexity classes and Kolmogorov complexity given in [2].

**Definition 2.1.** Say that $n \in \mathbb{N}$ is *super* if $\log^* n$ is an integer; i.e., if $n$ is in the range of the function $f : \mathbb{N} \to \mathbb{N}$ defined by the recursion $f(0) = 0$, and $f(n) = 2^{f(n-1)}$. $T \subset \{0\}^*$ is a *supertally set* if, for all $x \in T$, $|x|$ is super.

For example, the largest supertally set is $\widehat{T} = \{\lambda, 0, 0^2, 0^4, 0^{16}, 0^{65536}, 0^{2^{65536}}, 0^{2^{2^{65536}}}, \ldots\}$. The supertally sets are precisely the subsets of $\widehat{T}$. The property of supertally sets that will be useful is that, for any $n, m \in \mathbb{N}$ that are both super, if $n < m$, then $n \leq \log m$.

**Definition 2.2.** Let $U$ denote a universal Turing machine. For all $t : \mathbb{N} \to \mathbb{N}$ and $x \in \{0,1\}^*$, define $\mathrm{K}^t(x) = \min_{\pi \in \{0,1\}^*} \{\ |\pi|\ |\ U(\pi) = x \text{ in } t(|x|) \text{ steps }\}$ to be the *$t$-time-bounded Kolmogorov complexity* of $x$.

---

*University of Western Ontario, London, Ontario, Canada, ddoty@csd.uwo.ca

**Definition 2.3.** For all $l, t : \mathbb{N} \to \mathbb{N}$, define $K[l, t] = \{ x \in \{0, 1\}^* \mid K^t(x) \leq l(|x|) \}$.

In other words, $K[l, t]$ is the set of strings $x$ that cannot be computed in time $t(|x|)$ from a program of length at most $l(|x|)$. Note that, as is usually the case when dealing with Kolmogorov complexity, both of the bounding functions are in terms of the length of the *output* of the program.

Lipton [6] conjectured the following.

**Conjecture 2.4.** *For any $\epsilon > 0$ and time bound $t$, $\mathsf{NTIME}(t(n)) \subseteq \mathsf{DTIME}(2^{\epsilon t(n)})$.*

Lipton [5] also posed the following question, an negative answer to which would be stronger than Conjecture 2.4.

**Question 2.5.** *Is there a constant $c < 2$ such that, for all time bounds $t$, every nondeterministic Turing machine running in time $t$ can be simulated by a deterministic Turing machine running in time $c^{t(n)}$?*

The next theorem implies that non-relativizing techniques are required to give an affirmative answer to Question 2.5. The proof is based on Hartmanis' oracle [3] relative to which $\mathsf{P}$ is not $\mathsf{NP}$, constructed via Kolmogorov complexity (the first proof in [1] used diagonalization). Note that Conjecture 2.4 is also false relative to this oracle. The oracle constructed (though devised independently) appears to be similar to one constructed in Li and Vitanyi's textbook [4, Theorem 7.3.3], which also strengthens Hartmanis' technique, but for a different purpose, that of constructing an exponentially low set (an $A$ such that $\mathsf{E}^A = \mathsf{E}$).

**Theorem 2.6.** *There is an oracle $A$ and language $T$ such that $T \in \mathsf{NTIME}^A(n)$ and, for all $\delta > 0$, $T \notin \mathsf{DTIME}^A(2^{(1-\delta)n})$.*

*Proof.* Let $T \in \mathsf{DTIME}(2^{2n}) - \mathsf{DTIME}(2^n)$ be a supertally set. It is routine to construct such a set as in the proof of the time hierarchy theorem.

Define the oracle $A$ as follows. For each $n \in \mathbb{N}$, if $0^n \notin T$, then $A$ has no string of length $n$. If $0^n \in T$, then $A$ contains exactly one string $x$ of length $n$, chosen to be the first such string not in $K[n - 1, 2^n]$. In other words, $x$ is the first element of $\{0, 1\}^n$ that cannot be computed in time at most $2^n$ from a program of length at most $n - 1$. The abundance of incompressible strings ensures that such a string exists for all sufficiently large $n$.

Observe that $A \in \mathsf{DTIME}(2^{2n})$. To see why, let $x \in \{0, 1\}^n$ be the string whose membership in $A$ is to be decided. If $0^n \notin T$, which can be checked in time $2^{2n}$ by our choice of $T$, then $x \notin A$. If $0^n \in T$, then we enumerate all programs of length at most $n - 1$ (of which there are at most $2^n$), and simulate each of them for $2^n$ steps. These simulations take at most $2^{2n}$ steps. Then, given that $0^n \in T$, $x \in A$ if and only if $x$ is the first element of $\{0, 1\}^n$ that is not output by one of these programs. This shows $A \in \mathsf{DTIME}(2^{2n})$.

$T \in \mathsf{NTIME}^A(n)$ because $x \in T$ if and only if $x = 0^n$ for some $n \in \mathbb{N}$ and there exists $y \in \{0, 1\}^n$ such that $y \in A$. To complete the proof, let $\delta > 0$ and suppose for the sake of contradiction that $T \in \mathsf{DTIME}^A(2^{(1-\delta)n})$, via an oracle Turing machine $M^A$ running in time $2^{(1-\delta)n}$. We will show that this implies $T \in \mathsf{DTIME}(2^{(1-\delta)n}n^2)$, which contradicts our choice of $T \notin \mathsf{DTIME}(2^n)$, by showing how to eliminate the queries to $A$ during the execution of $M^A$.

Let $n \in \mathbb{N}$ and let $q \in \{0, 1\}^*$ be a string queried by $M^A$ on input $0^n$, where $n$ is super (we may assume $M$ immediately rejects all inputs not of this form). If $|q|$ is not super – which can be checked in time $|q| \leq 2^{(1-\delta)n}$ – then $q \notin A$. So we first check whether $|q|$ is super, and we know

that if not, then $q \notin A$. Since at most $2^{(1-\delta)n}$ such queries can be made by $M$, the totality of all such checks will take time at most $2^{(1-\delta)n}$.

Otherwise, if $|q|$ is super and $|q| < n$, then $|q| \leq \log n$, since $n$ is also super. Therefore, we can run the decider for $A$ to answer the query on $q$ in time at most $2^{2|q|} \leq n^2$. Since at most $2^{(1-\delta)n}$ such queries can be made by $M$, the totality of all such executions of the decider for $A$ will take time at most $2^{(1-\delta)n}n^2$.

Finally, suppose $|q|$ is super and $|q| \geq n$. We claim that $q \notin A$, implying that this query can also be eliminated and therefore $T \in \mathsf{DTIME}(2^{(1-\delta)n}n^2)$. Otherwise, suppose for the sake of contradiction that $q \in A$, and hence $q \notin \mathrm{K}[n-1, 2^n]$. Let $s_0, s_1, \ldots$ denote the standard enumeration of $\{0,1\}^*$. Then from the strings

- $M$ (of constant length),
- $s_n$ (of length $\log n$, from which $M$'s input $0^n$ can be produced), and
- $s_i$ ($i \in \mathbb{N}$ representing the order in which $q$ is first queried by $M$, out of all queries of length at least $n$; then $i \leq 2^{(1-\delta)n}$ because of $M$'s running time, implying $|s_i| \leq (1-\delta)n$), we can build a program of length $(1-\delta)n + \log n + O(1)$ that outputs $q$ in time $2^{(1-\delta)n}n^2$. This program executes $M^A(0^n)$ until it makes the $i^{\text{th}}$ query on a string of length at least $n$ and outputs that string. The first $i-1$ queries of length at least $n$ simply return "no" since $q$ is the only string of length at least $n$ in $A$ that is small enough to compute in time $2^{(1-\delta)n}$ (since the next largest string in $A$ has length at least $2^n$), and queries of length less than $n$ are handled as described above by using $A$'s decider if $|q|$ is super, and simply answering "no" otherwise. For sufficiently large $n$, this contradicts the fact that $q \notin \mathrm{K}[n-1, 2^n]$. □

**Corollary 2.7.** *There is an oracle $A$ such that, for all $c < 2$, $\mathsf{NTIME}^A(n) \not\subseteq \mathsf{DTIME}^A(c^n)$.*